 \documentclass[twocolumn,showpacs,preprintnumbers,amsmath,amssymb]{revtex4}

\usepackage{graphicx}
\usepackage{graphics,epsfig,psfig}

\begin{document}


\title{ An invariant distribution in static granular media}


\author{ T. Aste, T. Di Matteo, M. Saadatfar, T.J. Senden}
\address{ Department of Applied Mathematics, The Australian National University, 0200 Canberra, ACT, Australia}

 \author{M. Schr\"oter and H. L. Swinney  }
\address{Center for Nonlinear Dynamics and Department of Physics,
University of Texas at Austin, Austin, Texas 78712 USA }



\begin{abstract}

We have discovered an invariant distribution for local packing
configurations in static granular media.  This distribution holds in
experiments for packing fractions covering most of the range from
random loose packed to random close packed, for beads packed both in
air and in water.  Assuming only that there exist elementary cells in
which the system volume is subdivided, we derive from statistical
mechanics a distribution that is in accord with the observations. This
universal distribution function for granular media is analogous to the
Maxwell-Boltzmann distribution for molecular gasses.  

\end{abstract}

\maketitle

Granular materials are complex systems characterized by unusual static and dynamic properties.  
These systems are comprised of large numbers of dissipative macroscopic particles  
assembled into disordered structures.
The microscopic description of the system-state requires a very large number of
variables. However, there is a very large number of microscopic
configurations corresponding to the same macroscopic properties.
Edwards and coauthors \cite{Edwards89,Mehta89} have proposed that the
complexity of static granular systems could be disentangled by means
of a statistical mechanics approach reducing the description of the
system state to a few parameters only \cite{Fierro02,Makse02,
Behringer02,DAnna03,Ojha04,Richard05,ChamarraPRL06,Lechenault06,Metzger04,Metzger05}.  
An
essential part of Edwards' idea is that in static granular media
volume plays the role held by energy in usual thermodynamics.
Therefore, an understanding of the volume distribution function is the key
to connect microscopic details of the system with macroscopic state
variables.

Since granular materials are dissipative, they can change their static
configurations only when energy is injected into the system.  The
general idea underlying a statistical mechanics description is that
the properties of the system do not depend on the kind of energy
injections, but only on the portion of the configurational space that
the system explores under such action.  For instance, some classical
experiments~\cite{Knight95,Nowak97,Nowak98} obtained different
average packing fractions by tapping the container with different
intensities and different numbers of times. Similarly, in an
experiment by Schr\"oter et al.  \cite{Schroder05}, reproducible
average packing fractions were obtained by driving the system with
periodic trains of flow pulses in a fluidized bed.  More generally,
various controlled perturbations (tapping, rotating, pouring, etc.)
can produce packings with characteristic average packing fractions.
Within a statistical mechanics framework, if the system volume
($V_T$) is the relevant state variable, the system properties should
depend only on the achieved packing fraction and not on the 
preparation history.  
This hypothesis is examined in this paper using the
largest sets of experimental data on particle positions presently available.  

{\it Experiments.-} 
We analyze the structural properties of static granular packings
produced in 18 different experiments, 6 with acrylic spheres in air
and 12 with glass beads in water. The packing fractions $\rho$ range
from 0.56 to 0.64.  Three-dimensional density maps have been obtained
for these systems using X-ray Computed Tomography
\cite{AsteKioloa}. Coordinates of the bead centers have been
calculated for more than two million spheres with a precision better
than 1\% of their diameters, which is better than the 
uncertainty arising from polydispersity ($5\%$ for the glass particles
and $2\%$ for the acrylic particles). 

The 12 samples of glass beads were prepared using the fluidized bed
technique described in \cite{Schroder05}; each sample consisted of
about 145,000 spherical grains with diameters $250 \pm 13 \mu$m placed
in a cylindrical glass container with inner diameter $12.7$ mm.  The
beads were fluidized with pulses of deionized water with the flow rate
pulsed by a computer-controlled syringe pump.  During each flow pulse
the bed expanded until its height reached a stable value.  After each
flow pulse the bed was allowed to settle into a mechanically stable
configuration.  Packing fractions in the range between 0.56 and 0.60
were obtained by tuning the water flux.  The particle configurations
have been studied for sub-sets of grains that were at a distance
larger than four sphere diameters from the sample boundaries; there
were about 90,000 grains per sample.  

The 6 samples with acrylic spheres were described in \cite{AstePRE05}.
In particular, two samples contained 140,000 particles (90,000 in the
internal volume) with diameter $1.59 \pm 0.02$ mm in a cylindrical
container with an inner diameter of 55 mm, filled to a height of 75
mm.  The other four samples each had about 35,000 (15,000 in the
internal volume) acrylic spheres with diameter $1.00 \pm 0.02$ mm in the
same cylindrical container.

\begin{figure}
\centering
{\includegraphics[width=.5\textwidth]{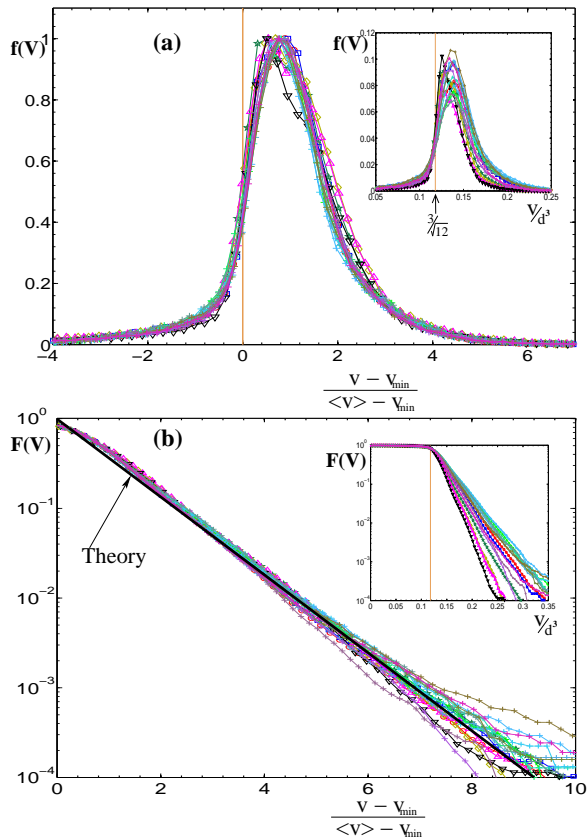}}
\caption{\footnotesize \label{f.dell} (a) Distributions of the
Delaunay cell volumes from 18 experiments collapse onto a
universal curve when plotted vs. $(V-V_{min})/(\left< V \right> -
V_{min})$.  The inset shows the same distributions plotted vs.
$V/d^3$. (b) The inverse cumulate distributions, $F(V) =
1-\int_{V_{min}}^V f(V) dV)$, also show evidence of a collapse onto
a universal curve when plotted vs. $(V-V_{min})/(\left< V
\right>- V_{min})$.  The line is the inverse cumulate
distributions for $f(V,k=1)$ (Eq.\ref{e.pV}).  The inset shows the
cumulate distributions plotted versus $V/d^3$.  The distributions are
obtained from a statistical analysis over more than six million cells
in 18 different experiments with three different kinds of beads in air
and in water.  The open symbols correspond to the 6 experiments with
with dry acrylic beads \cite{AstePRE05}: circles sample A ($\rho =
0.586$); squares sample B ($\rho = 0.596$); stars sample C ($\rho =
0.619$); diamonds sample D ($\rho = 0.626$); triangles up sample E
($\rho = 0.630$); triangles down sample F ($\rho = 0.64$).  The `+'
corresponds to the 12 experiments with packing fractions $0.56 \le
\rho \le 0.60$ made with glass spheres in of fluidized bed.  }
\end{figure}

{\it Theory.-}  Since the original work by Edwards and co-authors, several statistical
mechanics approaches have been proposed to describe granular materials
\cite{Edwards89,Mehta89,Makse02,
Behringer02,DAnna03,Ojha04,Richard05,Ciamarra06,Lechenault06}.  We
make here the minimal number of assumptions in considering a granular
system at rest that occupies a given volume $V_T$.  We assume 
that the system can be subdivided into $C$ elementary cells with volumes
$v_i$ ($i=1...C$) with $\sum_{i=1}^C v_i = V_T$.  Under the sole
assumption that any assembly of such elementary volumes will produce
proper mechanically stable packings, we obtain the probability
distribution for the cell-elementary volumes:
\begin{equation}
p(v) = \frac{1}{\chi}e^{-({v-v_{min}})/{\chi}}\;\;,
\label{e.pV}
\end{equation}
with 
\begin{equation}
\chi = \left< v \right> - v_{min}  \;\;,
\label{e.pVchi}
\end{equation}
where $ \left< v \right> = V_T/C$ is the average volume per elementary
cell, $v_{min}$ is the minimum volume attainable by the elementary
cells and $\chi$ is an intensive thermodynamic parameter accounting
for the exchange of volume between the elementary cell and the
surrounding volume ``reservoir''.  There is no need to specify the
nature and kind of such ``elementary cells''; the only
assumptions are that they exist, that they can have any volume
larger than $v_{min}$, and that their assembly fills a volume $V_T$.  Space
can be divided arbitrarily into pieces. Common examples of such
partitions into space-filling blocks are the Delaunay and the
Vorono\"{\i} decompositions \cite{Voronoi,AstePRL06,ppp}.  Such cells
do not coincide with the elementary ones, but they might be assemblies
of such elementary cells.  For instance, in a Delaunay partition of a
three-dimensional packing there are about six times more cells than in
the Vorono\"{\i} decomposition.  The present theory applies to any
degree of space partition made by any agglomerate local structure made
of a given number $k$ of elementary cells.  From Eq.~\ref{e.pV} the
probability distribution for the volumes of such agglomerate of $k$
elementary cells is
\begin{equation}
f(V,k) = \frac{k^{k}}{(k-1)! } \frac{(V - V_{min})^{(k-1)} }{(\left< V
\right> - V_{min})^{k}} \exp \left( {-k \frac{V-V_{min}}{\left< V
\right> - V_{min} } } \right)\;\;\;,
\label{e.pVk}
\end{equation}
with $\left< V \right> = V_T/C$ and $V_{min} = k v_{min}$ being
respectively the average and the minimum volumes for a given
packing. It follows from Eqs.~\ref{e.pV} and \ref{e.pVk} that
 \begin{equation}
 \chi =  \frac{ \left< V \right> - V_{min}}{k} \;\;\;,
\label{e.chi}
\end{equation}
which is the average free-volume per elementary cell.  Therefore, the
intensive variable $\chi$ is a measure of the kind and the degree of
space-partition into elementary cells.

We have obtained $f(V,k)$ assuming that the cells are uniquely
characterized by their volumes, and that any combination of $C$ cells
with arbitrary volumes $v_i \ge v_{min}$ will produce a structurally
stable, space-filling system of cells.  This is possible in 
one-dimension only where, indeed, the packing is an arbitrary
disposition of grain centers at distances larger or equal than
$v_{min}$.  In this case, the elementary cells are the Delaunay cells
\cite{ppp}, which are the segments in between two successive grain
centers; the distribution of their sizes is exactly described by
$p(v)$.  In three dimensions, the Delaunay cells are tetrahedra with
vertices on the centers of neighborhood grains chosen in a way that no
other grains in the packing have centers within the circumsphere of
each Delaunay tetrahedron.  Clearly, in this case, the Delaunay cells
are not uniquely described by their volumes and an arbitrary
collection of cells is neither space-filling nor mechanically stable.

{\it Results.-}
We observe that the distributions of the Delaunay volumes obtained from the 18 sets of
data for different kinds of beads in different media and different
conditions are all described well by the {\it same} function, as shown
Fig.\ref{f.dell}(a).  The collapse of the data onto a single curve was
obtained using $V_{min}= \sqrt{2}/12 d^3$ (with $d$ the sphere
diameter), which is the volume of a regular Delaunay tetrahedron for
four spheres in contact (the smallest compact tetrahedron
\cite{AstePRL06}).  
The tail of the observed distribution function for
the Delauney volumes is exponential, as in Eq.~\ref{e.pV} with $k \sim
1$. 
The data fit well to the inverse cumulate
distribution, $F(V) = 1-\int_{v_{min}}^V f(v',1) dv'$, as
Fig.\ref{f.dell}(b) shows; the agreement extends over four orders of
magnitude and uses no adjustable parameters.  The agreement suggests
that Delaunay cells might be considered candidates for the
``elementary cells''; however, it was noted in \cite{AstePRL06} that, at small $V-V_{min}$ 
in the region where the spheres make contact ($V < \sqrt{3}d^3/12$),
the empirical distributions deviate from the simple exponential form
predicted by $p(v)$, which indicates that other constraints such as
mechanical equilibrium should be also taken into account to correctly
describe such region.

\begin{figure} 
\centering
{\includegraphics[width=0.5\textwidth]{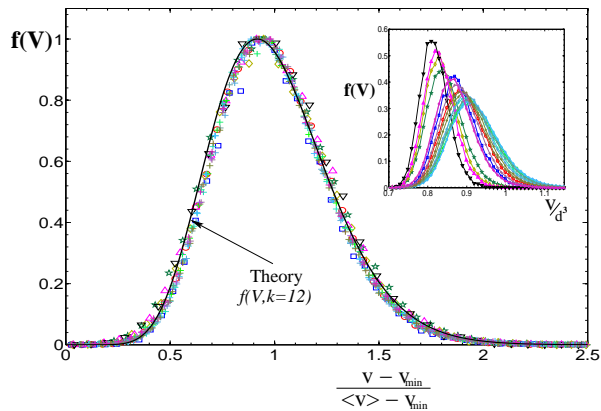}}
\caption{\footnotesize The distributions of the Vorono\"{\i} cell
volumes collapse onto a universal curve when plotted vs.
$(V-V_{min})/(\left< V \right>- V_{min})$.  The theoretical line is
$f(V,k=12)$ (Eq.~\ref{e.pV}).  The inset shows the distributions
plotted versus $V/d^3$.  The data and the symbols are the 
same as in Fig.~\ref{f.dell}.}
\label{f.vor}
\end{figure}

An alternative way for dividing space into space-filling cells is the
Vorono\"{\i} partition.  In one dimension, the Vorono\"{\i} cell
around grain `$i$' is constructed by taking the segment between the
two mid-points between grains $i-1$ and $i$, and between grains $i$
and $i+1$.  The size of such a segment is equal to $(v_i + v_{i+1})/2$
where $v_i$ and $v_{i+1}$ are the distances between the two
consecutive couples of points.  Therefore, the probability of finding
a one-dimensional Vorono\"{\i} cell of size $V$ is associated with the
probability of finding two successive Delaunay cells with sizes $v_1 +
v_2 = 2V$. This probability is $\int_{2 v_{min}}^{V-v_{min} }
p(v_1) p(v-v_1) dv_1 \propto (V- 2 v_{min}) \exp[-(V- 2
v_{min})/(\left< V \right>- 2v_{min})] \propto f(V,k=2)$, with $\left<
V \right>$ the average size of the Vorono\"{\i} cell.  In three
dimensions, the Vorono\"{\i} cell can be also seen as the combination
of several elementary cells with distribution $f(V,k)$; however, in
this case the number of sub-cells involved is not fixed at $k=2$ but
depends on the kind of packing.

We observe that data for over a million of Vorono\"{\i} cells from all 18 experiments
for dry and wet packings of glass and acrylic spheres collapse to the
the same distribution function (see Fig.~\ref{f.vor}).  In this
collapse of the data there no adjustable paramaters, just $\left< V
\right> = V_T/$~({\it number of grains}) and $V_{min} =
5^{(5/4)}/\sqrt{2(29+ 13 \sqrt{5}) } d^3 \simeq 0.694 d^3$, which is
the smallest Vorono\"{\i} cell that can be built in a equal-spheres
packing~\cite{ppp}.  Figure~\ref{f.vor} shows that such
universal distribution function is well described by Eq.~\ref{e.pV}
with $k=12$, which indicates that about 12 elementary cells contribute
in building each Vorono\"{\i} cell.  Such a number is
meaningful, since about 12 spheres are expected to be found in  the
close neighboring of any given sphere in the packing.

\begin{figure} 
\centering
{\includegraphics[width=0.35\textwidth]{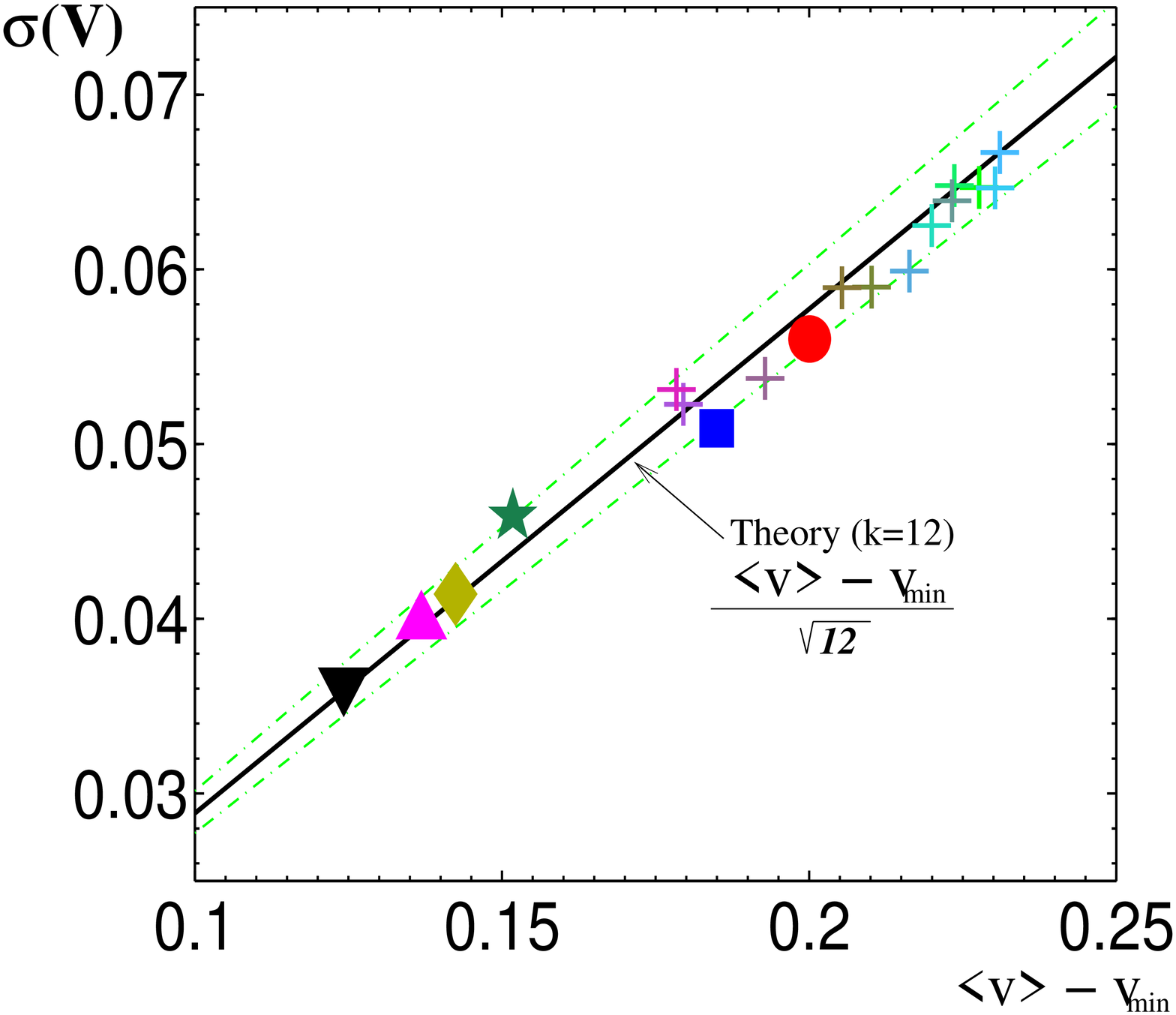}}
\caption{\footnotesize The standard deviations of the
Vorono\"{\i} volume distributions in the 18
experiments fit the linear relation given 
by Eq.~\ref{e.chi_sig} with $k = 12$. The dash line
above (below) the solid line corresponds to $k=11$ ($k=13$).}
\label{f.chi}
\end{figure}

A further demonstration that such statistical distributions are independent of the details of a
sample and the method of sample preparation is provided by the
behavior of the volume fluctuations, which can be calculated directly
from the relation (see Eq.~\ref{e.pV}),
 \begin{equation}
\sigma^2(V) = \left< (V - \left< V \right>)^2\right> = \chi^2
\frac{\partial \left< V \right>}{\partial \chi} \;\;\;,
\label{e.fluct}
\end{equation}
which becomes, using $ \partial \left< V \right>/\partial \chi = k $
(from Eq.~\ref{e.chi}),
 \begin{equation}
 \sigma(V) = \sqrt{k} \;\; \chi  =  \frac{\left< V \right> - V_{min}}{\sqrt{k}} \;\;\;.
\label{e.chi_sig}
\end{equation}

The good correspondence of the data from the Vorono\"{\i} volume distributions
with Eq.~\ref{e.chi_sig} (Fig.~\ref{f.chi}) provides further evidence that 
$\left< V \right> $ and $k$ are the relevant control parameters, independent
of the system and its preparation.

{\it Conclusion.-}
We have shown that for a broad set of granular packings
of different monodisperse spherical grains, prepared by
using different procedures, the local volumes are described by
a universal distribution function (Eq.~\ref{e.pVk}), which can be predicted from a
minimal set of assumptions.  The only tunable parameter in the present
theory is $k$, which has the value 12 for 
Vorono\"{\i} decompositions throughout the
accessible density ranges of the different static granular packings studied.  
Interestingly, for granular gasses the same empirical distribution (Eq.~\ref{e.pVk})
applies but with $k = 5.586$ \cite{Gilbert62,Pineda04}.
A similar kind of distribution (Gamma distribution) with different values of the parameter $k$ have also been observed in two dimensional Vorono\"{\i} tessellation generated from  disk packings \cite{Kumar05}.
Therefore, the parameter $k$ could be the `structure parameter' which
depends on the system-phase.  
Within the framework of a statistical
mechanics description, Eq.~\ref{e.pVk} can be regarded as the
analogous to the Maxwell-Boltzmann distribution for granular media.

We thank  Mario Nicodemi,
Massimo Pica Ciamarra, Tom 
Truskett and Philip Metzger for helpful discussions and comments.  
This work was partially supported by the ARC discovery project DP0450292. MS and HLS
were supported by the Robert A. Welch Foundation.


\begin{thebibliography}{25}
\expandafter\ifx\csname natexlab\endcsname\relax\def\natexlab#1{#1}\fi
\expandafter\ifx\csname bibnamefont\endcsname\relax
  \def\bibnamefont#1{#1}\fi
\expandafter\ifx\csname bibfnamefont\endcsname\relax
  \def\bibfnamefont#1{#1}\fi
\expandafter\ifx\csname citenamefont\endcsname\relax
  \def\citenamefont#1{#1}\fi
\expandafter\ifx\csname url\endcsname\relax
  \def\url#1{\texttt{#1}}\fi
\expandafter\ifx\csname urlprefix\endcsname\relax\def\urlprefix{URL }\fi
\providecommand{\bibinfo}[2]{#2}
\providecommand{\eprint}[2][]{\url{#2}}

\bibitem[{\citenamefont{Edwards and Oakeshott}(1989)}]{Edwards89}
\bibinfo{author}{\bibfnamefont{S.}~\bibnamefont{Edwards}} \bibnamefont{and}
  \bibinfo{author}{\bibfnamefont{R.}~\bibnamefont{Oakeshott}},
  \bibinfo{journal}{Physica A} \textbf{\bibinfo{volume}{157}},
  \bibinfo{pages}{1080} (\bibinfo{year}{1989}).

\bibitem[{\citenamefont{Mehta and Edwards}(1989)}]{Mehta89}
\bibinfo{author}{\bibfnamefont{A.}~\bibnamefont{Mehta}} \bibnamefont{and}
  \bibinfo{author}{\bibfnamefont{S.~F.} \bibnamefont{Edwards}},
  \bibinfo{journal}{Physica A} \textbf{\bibinfo{volume}{157}},
  \bibinfo{pages}{1091} (\bibinfo{year}{1989}).

\bibitem[{\citenamefont{Fierro et~al.}(2002)\citenamefont{Fierro, Nicodemi, and
  Coniglio}}]{Fierro02}
\bibinfo{author}{\bibfnamefont{A.}~\bibnamefont{Fierro}},
  \bibinfo{author}{\bibfnamefont{M.}~\bibnamefont{Nicodemi}}, \bibnamefont{and}
  \bibinfo{author}{\bibfnamefont{A.}~\bibnamefont{Coniglio}},
  \bibinfo{journal}{Europhys. Lett.} \textbf{\bibinfo{volume}{59}},
  \bibinfo{pages}{642} (\bibinfo{year}{2002}).

\bibitem[{\citenamefont{Makse and Kurchan}(2002)}]{Makse02}
\bibinfo{author}{\bibfnamefont{H.~A.} \bibnamefont{Makse}} \bibnamefont{and}
  \bibinfo{author}{\bibfnamefont{J.}~\bibnamefont{Kurchan}},
  \bibinfo{journal}{Nature} \textbf{\bibinfo{volume}{415}},
  \bibinfo{pages}{614} (\bibinfo{year}{2002}).

\bibitem[{\citenamefont{Behringer}(2002)}]{Behringer02}
\bibinfo{author}{\bibfnamefont{B.}~\bibnamefont{Behringer}},
  \bibinfo{journal}{Nature} \textbf{\bibinfo{volume}{415}},
  \bibinfo{pages}{594} (\bibinfo{year}{2002}).

\bibitem[{\citenamefont{D'Anna et~al.}(2003)\citenamefont{D'Anna, Mayor,
  Barrat, Loreto, and Nori}}]{DAnna03}
\bibinfo{author}{\bibfnamefont{G.}~\bibnamefont{D'Anna}},
  \bibinfo{author}{\bibfnamefont{P.}~\bibnamefont{Mayor}},
  \bibinfo{author}{\bibfnamefont{A.}~\bibnamefont{Barrat}},
  \bibinfo{author}{\bibfnamefont{V.}~\bibnamefont{Loreto}}, \bibnamefont{and}
  \bibinfo{author}{\bibfnamefont{F.}~\bibnamefont{Nori}},
  \bibinfo{journal}{Nature} \textbf{\bibinfo{volume}{424}},
  \bibinfo{pages}{909} (\bibinfo{year}{2003}).

\bibitem[{\citenamefont{Ojha et~al.}(2004)\citenamefont{Ojha, Lemieux, Dixon,
  Liu, and Durian}}]{Ojha04}
\bibinfo{author}{\bibfnamefont{R.~P.} \bibnamefont{Ojha}},
  \bibinfo{author}{\bibfnamefont{P.~A.} \bibnamefont{Lemieux}},
  \bibinfo{author}{\bibfnamefont{P.~K.} \bibnamefont{Dixon}},
  \bibinfo{author}{\bibfnamefont{A.~J.} \bibnamefont{Liu}}, \bibnamefont{and}
  \bibinfo{author}{\bibfnamefont{D.~J.} \bibnamefont{Durian}},
  \bibinfo{journal}{Nature} \textbf{\bibinfo{volume}{427}},
  \bibinfo{pages}{521} (\bibinfo{year}{2004}).

\bibitem[{\citenamefont{Richard et~al.}(2005)\citenamefont{Richard, Nicodemi,
  Delannay, Ribiere, and Bideau}}]{Richard05}
\bibinfo{author}{\bibfnamefont{P.}~\bibnamefont{Richard}},
  \bibinfo{author}{\bibfnamefont{M.}~\bibnamefont{Nicodemi}},
  \bibinfo{author}{\bibfnamefont{R.}~\bibnamefont{Delannay}},
  \bibinfo{author}{\bibfnamefont{P.}~\bibnamefont{Ribiere}}, \bibnamefont{and}
  \bibinfo{author}{\bibfnamefont{D.}~\bibnamefont{Bideau}},
  \bibinfo{journal}{Nature Materials} \textbf{\bibinfo{volume}{4}},
  \bibinfo{pages}{121} (\bibinfo{year}{2005}).

\bibitem[{\citenamefont{Ciamarra
  et~al.}(2006{\natexlab{a}})\citenamefont{Ciamarra, Nicodemi, and
  Coniglio}}]{ChamarraPRL06}
\bibinfo{author}{\bibfnamefont{M.~P.} \bibnamefont{Ciamarra}},
  \bibinfo{author}{\bibfnamefont{M.}~\bibnamefont{Nicodemi}}, \bibnamefont{and}
  \bibinfo{author}{\bibfnamefont{A.}~\bibnamefont{Coniglio}},
  \bibinfo{journal}{Phys. Rev. Lett.}  \textbf{\bibinfo{volume}{97}},\bibinfo{pages}{158001}
  (\bibinfo{year}{2006}{\natexlab{a}}).

\bibitem[{\citenamefont{Lechenault et~al.}(2006)\citenamefont{Lechenault,
  da~Cruz, Dauchot, and Bertin}}]{Lechenault06}
\bibinfo{author}{\bibfnamefont{F.}~\bibnamefont{Lechenault}},
  \bibinfo{author}{\bibfnamefont{F.}~\bibnamefont{da~Cruz}},
  \bibinfo{author}{\bibfnamefont{O.}~\bibnamefont{Dauchot}}, \bibnamefont{and}
  \bibinfo{author}{\bibfnamefont{E.}~\bibnamefont{Bertin}},
  \bibinfo{journal}{J. Stat. Mech.} \textbf{\bibinfo{volume}{P07009}},
  \bibinfo{pages}{1742} (\bibinfo{year}{2006}).

\bibitem[{\citenamefont{Metzger}(2004)}]{Metzger04}
\bibinfo{author}{\bibfnamefont{P.~T.} \bibnamefont{Metzger}},
  \bibinfo{journal}{Phys. Rev. E.} \textbf{\bibinfo{volume}{70}},\bibinfo{pages}{051303}
  (\bibinfo{year}{2004}).

\bibitem[{\citenamefont{Metzger and Donahue}(2005)}]{Metzger05}
\bibinfo{author}{\bibfnamefont{P.~T.} \bibnamefont{Metzger}} \bibnamefont{and}
  \bibinfo{author}{\bibfnamefont{C.~M.} \bibnamefont{Donahue}},
  \bibinfo{journal}{Phys. Rev. Lett.} \textbf{\bibinfo{volume}{94}},
  \bibinfo{pages}{148001} (\bibinfo{year}{2005}).

\bibitem[{\citenamefont{Knight et~al.}(1995)\citenamefont{Knight, Fandrich,
  Lau, Jeager, and Nagel}}]{Knight95}
\bibinfo{author}{\bibfnamefont{J.~B.} \bibnamefont{Knight}},
  \bibinfo{author}{\bibfnamefont{C.~G.} \bibnamefont{Fandrich}},
  \bibinfo{author}{\bibfnamefont{C.~N.} \bibnamefont{Lau}},
  \bibinfo{author}{\bibfnamefont{H.~M.} \bibnamefont{Jeager}},
  \bibnamefont{and} \bibinfo{author}{\bibfnamefont{S.~R.} \bibnamefont{Nagel}},
  \bibinfo{journal}{Phys Rev E} \textbf{\bibinfo{volume}{51}},
  \bibinfo{pages}{3957} (\bibinfo{year}{1995}).

\bibitem[{\citenamefont{Nowak et~al.}(1997)\citenamefont{Nowak, Knight,
  Povinelli, Jeager, and Nagel}}]{Nowak97}
\bibinfo{author}{\bibfnamefont{E.~R.} \bibnamefont{Nowak}},
  \bibinfo{author}{\bibfnamefont{J.~B.} \bibnamefont{Knight}},
  \bibinfo{author}{\bibfnamefont{M.~L.} \bibnamefont{Povinelli}},
  \bibinfo{author}{\bibfnamefont{H.~M.} \bibnamefont{Jeager}},
  \bibnamefont{and} \bibinfo{author}{\bibfnamefont{S.~R.} \bibnamefont{Nagel}},
  \bibinfo{journal}{Powder Technol.} \textbf{\bibinfo{volume}{94}},
  \bibinfo{pages}{79} (\bibinfo{year}{1997}).

\bibitem[{\citenamefont{Nowak et~al.}(1998)\citenamefont{Nowak, Knight,
  BenNaim, Jeager, and Nagel}}]{Nowak98}
\bibinfo{author}{\bibfnamefont{E.~R.} \bibnamefont{Nowak}},
  \bibinfo{author}{\bibfnamefont{J.~B.} \bibnamefont{Knight}},
  \bibinfo{author}{\bibfnamefont{E.}~\bibnamefont{BenNaim}},
  \bibinfo{author}{\bibfnamefont{H.~M.} \bibnamefont{Jeager}},
  \bibnamefont{and} \bibinfo{author}{\bibfnamefont{S.~R.} \bibnamefont{Nagel}},
  \bibinfo{journal}{Phys. Rev. E} \textbf{\bibinfo{volume}{57}},
  \bibinfo{pages}{1971} (\bibinfo{year}{1998}).

\bibitem[{\citenamefont{Schr\"oter et~al.}(2005)\citenamefont{Schr\"oter,
  Goldman, and Swinney}}]{Schroder05}
\bibinfo{author}{\bibfnamefont{M.}~\bibnamefont{Schr\"oter}},
  \bibinfo{author}{\bibfnamefont{D.~I.} \bibnamefont{Goldman}},
  \bibnamefont{and} \bibinfo{author}{\bibfnamefont{H.~L.}
  \bibnamefont{Swinney}}, \bibinfo{journal}{Phys. Rev. E.}
  \textbf{\bibinfo{volume}{71}}, \bibinfo{pages}{30301 (R)}
  (\bibinfo{year}{2005}).

\bibitem[{\citenamefont{Aste et~al.}(2004)\citenamefont{Aste, Saadatfar,
  Sakellariou, and Senden}}]{AsteKioloa}
\bibinfo{author}{\bibfnamefont{T.}~\bibnamefont{Aste}},
  \bibinfo{author}{\bibfnamefont{M.}~\bibnamefont{Saadatfar}},
  \bibinfo{author}{\bibfnamefont{A.}~\bibnamefont{Sakellariou}},
  \bibnamefont{and} \bibinfo{author}{\bibfnamefont{T.}~\bibnamefont{Senden}},
  \bibinfo{journal}{Physica A} \textbf{\bibinfo{volume}{339}},
  \bibinfo{pages}{16} (\bibinfo{year}{2004}).

\bibitem[{\citenamefont{Aste et~al.}(2005)\citenamefont{Aste, Saadatfar, and
  Senden}}]{AstePRE05}
\bibinfo{author}{\bibfnamefont{T.}~\bibnamefont{Aste}},
  \bibinfo{author}{\bibfnamefont{M.}~\bibnamefont{Saadatfar}},
  \bibnamefont{and} \bibinfo{author}{\bibfnamefont{T.~J.}
  \bibnamefont{Senden}}, \bibinfo{journal}{Phys. Rev. E.}
  \textbf{\bibinfo{volume}{71}}, \bibinfo{pages}{061302}
  (\bibinfo{year}{2005}).

\bibitem[{\citenamefont{Ciamarra
  et~al.}(2006{\natexlab{b}})\citenamefont{Ciamarra, Nicodemi, and
  Coniglio}}]{Ciamarra06}
\bibinfo{author}{\bibfnamefont{M.~P.} \bibnamefont{Ciamarra}},
  \bibinfo{author}{\bibfnamefont{M.}~\bibnamefont{Nicodemi}}, \bibnamefont{and}
  \bibinfo{author}{\bibfnamefont{A.}~\bibnamefont{Coniglio}},
  \bibinfo{journal}{preprint.}  (\bibinfo{year}{2006}{\natexlab{b}}).

\bibitem[{\citenamefont{Vorono\"{\i}}(1908)}]{Voronoi}
\bibinfo{author}{\bibfnamefont{G.}~\bibnamefont{Vorono\"{\i}}},
  \bibinfo{journal}{J. reine angew. Math.} \textbf{\bibinfo{volume}{134}},
  \bibinfo{pages}{198} (\bibinfo{year}{1908}).

\bibitem[{\citenamefont{Aste}(2006)}]{AstePRL06}
\bibinfo{author}{\bibfnamefont{T.}~\bibnamefont{Aste}}, \bibinfo{journal}{Phys.
  Rev. Lett.} \textbf{\bibinfo{volume}{96}}, \bibinfo{pages}{018002}
  (\bibinfo{year}{2006}).

\bibitem[{\citenamefont{Aste and Weaire}(2000)}]{ppp}
\bibinfo{author}{\bibfnamefont{T.}~\bibnamefont{Aste}} \bibnamefont{and}
  \bibinfo{author}{\bibfnamefont{D.}~\bibnamefont{Weaire}},
  \emph{\bibinfo{title}{The Pursuit of Perfect Packing}}
  (\bibinfo{publisher}{Institute of Physics, Bristol}, \bibinfo{year}{2000}).

\bibitem[{\citenamefont{Gilbert}(1962)}]{Gilbert62}
\bibinfo{author}{\bibfnamefont{E.~N.} \bibnamefont{Gilbert}},
  \bibinfo{journal}{Ann. Math. Stat.} \textbf{\bibinfo{volume}{33}},
  \bibinfo{pages}{958} (\bibinfo{year}{1962}).

\bibitem[{\citenamefont{Pineda et~al.}(2004)\citenamefont{Pineda, Bruna, and
  Crespo}}]{Pineda04}
\bibinfo{author}{\bibfnamefont{E.}~\bibnamefont{Pineda}},
  \bibinfo{author}{\bibfnamefont{P.}~\bibnamefont{Bruna}}, \bibnamefont{and}
  \bibinfo{author}{\bibfnamefont{D.}~\bibnamefont{Crespo}},
  \bibinfo{journal}{Phys. Rev. E.} \textbf{\bibinfo{volume}{70}},
  \bibinfo{pages}{066119} (\bibinfo{year}{2004}).

\bibitem[{\citenamefont{Kumar and Kumaran}(2005)}]{Kumar05}
\bibinfo{author}{\bibfnamefont{V.~S.} \bibnamefont{Kumar}} \bibnamefont{and}
  \bibinfo{author}{\bibfnamefont{V.}~\bibnamefont{Kumaran}},
  \bibinfo{journal}{J. Chem. Phys.} \textbf{\bibinfo{volume}{123}},
  \bibinfo{pages}{114501} (\bibinfo{year}{2005}).

\end{thebibliography}
\end{document}